\documentclass[12pt]{article}
\usepackage{epsfig}

\newlength{\abstwidth}
{\end{list}}
\newcounter{enumct}

\begin{document}

\sloppy

\begin{flushright}
LU TP 00--53\\
hep-ph/0012187\\
December 2000
\end{flushright}

\vspace{2cm}

\begin{center}
{\LARGE\bf QCD Radiation off Heavy Particles%
\footnote{to appear in the Proceedings of the Linear Collider %
Workshop 2000, Fermilab, October 24--28, 2000}}\\[10mm]
{\Large T. Sj\"ostrand\footnote{torbjorn@thep.lu.se}} \\[3mm]
{\it Department of Theoretical Physics,}\\[1mm]
{\it Lund University,}\\[1mm]
{\it S\"olvegatan 14A,}\\[1mm]
{\it S-223 62 Lund, Sweden}
\end{center}
 
\vspace{2cm}
 
\begin{center}
{\bf Abstract}\\[2ex]
\begin{minipage}{\abstwidth}
An algorithm for an improved description of final-state QCD radiation
is introduced. It is matched to the first-order matrix elements for 
gluon emission in a host of decays, for processes within
the Standard Model and the Minimal Supersymmetric extension thereof.
\end{minipage}
\end{center}
 
\vspace{1cm}

\noindent 
\rule{160mm}{0.5mm}

\vspace{1cm}

The objective of this article is to summarize the improved description
of parton shower evolution \cite{newshow} recently introduced 
starting with \textsc{Pythia} 6.154 \cite{pythia}. In particular, 
process-specific $\mathcal{O}(\alpha_{\mathrm{s}})$ matrix elements
for gluon emission in decays $a \to bc$ are used to match the shower
description to the correct emission rate in the hard-gluon region,
and to provide the proper amount of `dead cone' \cite{deadcone}
suppression of collinear gluon emission off massive particles. The 
original motivation was to improve the understanding of $b\overline{b}$ 
events at LEP1. For linear colliders the applications to top, Higgs and 
SUSY physics are very important.

The traditional final-state shower algorithm \cite{oldshow} 
in \textsc{Pythia} is based on an evolution in $Q^2 = m^2$, i.e. potential
branchings are considered in order of decreasing mass. A branching
$d \to ef$ is then characterized by $m_d^2$ and $z = E_e/E_d$.
For the process $\gamma^*/Z \to q\overline{q}$, the first gluon emission 
off both $q$ and $\overline{q}$ are corrected to the first-order matrix 
elements for $\gamma^*/Z \to q\overline{q} g$. (The $\alpha_{\mathrm{s}}$ 
and the Sudakov form factor are omitted from the comparison, since the 
shower procedure here attempts to include higher-order effects absent in 
the first-order matrix elements.)

This matching is well-defined for massless quarks, and was originally
used unchanged for massive ones. A first attempt to include massive 
matrix elements did not compensate for mass effects in the shower
kinematics, and therefore came to exaggerate the suppression of 
radiation off heavy quarks \cite{QCDWG}. Now the shower has 
been modified to solve this issue, and also improved and extended 
better to cover a host of different reactions \cite{newshow}. 

The starting point is the calculation of processes
$a \to bc$ and $a \to bcg$, where the ratio
\begin{equation}
W_{\mathrm{ME}}(x_1,x_2) =
\frac{1}{\sigma(a \to bc)} \, 
\frac{{\mathrm{d}}\sigma(a \to bcg)}{{\mathrm{d}} x_1 \, {\mathrm{d}} x_2}
\label{WME}
\end{equation}
gives the process-dependent differential gluon-emission rate. 
Here the phase space variables are $x_1 = 2E_b/m_a$ and  
$x_2 = 2E_c/m_a$, expressed in the rest frame of parton $a$.
Using the standard model and the minimal supersymmetric extension
thereof as templates, a wide selection of colour and spin structures
have been addressed, exemplified by $Z^0 \to q\overline{q}$,
$t \to b W^+$, $H^0 \to q\overline{q}$, $t \to b H^+$,
$Z^0 \to \tilde{q}\overline{\tilde{q}}$, $\tilde{q} \to \tilde{q}' W^+$, 
$H^0 \to \tilde{q}\overline{\tilde{q}}$, $\tilde{q} \to \tilde{q}' H^+$, 
$\tilde{\chi} \to q\overline{\tilde{q}}$, $\tilde{q} \to q\tilde{\chi}$,
$t \to \tilde{t}\tilde{\chi}$, $\tilde{g} \to q\overline{\tilde{q}}$, 
$\tilde{q} \to q\tilde{g}$, and $t \to \tilde{t}\tilde{g}$. 
The mass ratios $r_1 = m_b / m_a$ and $r_2 = m_c/m_a$ 
have been kept as free parameters. When allowed, processes have
been calculated for an arbitrary mixture of ``parities'', i.e.
without or with a $\gamma_5$ factor, like in the vector/axial vector 
structure of $\gamma^*/Z$. All the matrix elements are encoded in the new 
function \texttt{PYMAEL(NI,X1,X2,R1,R2,ALPHA)}, where \texttt{NI} 
distinguishes the matrix elements and \texttt{ALPHA} is related to the 
$\gamma_5$ admixture.

In order to match to the singularity structure of the massive matrix 
elements, the evolution variable $Q^2$ is changed from $m^2$ to 
$m^2 - m_{\mathrm{on-shell}}^2$, i.e. $1/Q^2$ is the propagator of a 
massive particle. Furthermore, the $z$ variable of a branching needs to 
be redefined, which is achieved by reducing the three-momenta
of the daughters in the rest frame of the mother. For the shower history 
$b \to bg$ this gives a differential probability
\begin{equation}
W_{\mathrm{PS,1}}(x_1,x_2) 
= \frac{\alpha_{\mathrm{s}}}{2\pi} \, C_F \, \frac{{\mathrm{d}} Q^2}{Q^2} 
\, \frac{2 \, {\mathrm{d}} z}{1-z} \, \frac{1}{{\mathrm{d}} x_1 \, 
{\mathrm{d}} x_2} = \frac{\alpha_{\mathrm{s}}}{2\pi} \, C_F \,
\frac{2}{x_3 \, (1 + r_2^2 - r_1^2 - x_2)}  ~,
\end{equation}
where the numerator $1 + z^2$ of the splitting kernel for $q \to q g$ 
has been replaced by a 2 in the shower algorithm. For a process with only 
one radiating parton in the final state, such as $t \to b W^+$, the 
ratio $W_{\mathrm{ME}}/W_{\mathrm{PS,1}}$ gives the acceptance probability 
for an emission in the shower. The singularity structure exactly agrees 
between ME and PS, giving a well-behaved ratio always below unity. If both 
$b$ and $c$ can radiate, there is a second possible shower history that 
has to be considered. The matrix element is here split in two parts, one 
arbitrarily associated with $b \to b g$ branchings and the other with 
$c \to c g$ ones. A convenient choice is 
$W_{\mathrm{ME,1}} = W_{\mathrm{ME}} (1 + r_1^2 - r_2^2 - x_1)/x_3$ and 
$W_{\mathrm{ME,2}} = W_{\mathrm{ME}} (1 + r_2^2 - r_1^2 - x_2)/x_3$,
which again gives matching singularity structures in 
$W_{{\mathrm{ME,}}i}/W_{{\mathrm{PS,}}i}$ and thus a
well-behaved Monte Carlo procedure. 

Also subsequent emissions of gluons off the primary particles are 
corrected to $W_{\mathrm{ME}}$. To this 
end, a reduced-energy system is constructed, which retains the 
kinematics of the branching under consideration but omits the gluons 
already emitted, so that an effective three-body shower state can be 
mapped to an $(x_1, x_2, r_1, r_2)$ set of variables. For light quarks 
this procedure is almost equivalent with the original one of using the  
simple universal splitting kernels after the first branching. For
heavy quarks it offers an improved modelling of mass effects also in 
the collinear region.

Some further changes have been introduced, a few minor as default and
some more significant ones as non-default options \cite{newshow}. 
This includes the description of coherence effects and 
$\alpha_{\mathrm{s}}$ arguments, in general and more specifically for 
secondary heavy flavour production by gluon splittings.

Further issues remain to be addressed, e.g. radiation off particles
with non-negligible width. In general, however, the new shower should 
allow an improved description of gluon radiation in many different
processes. Where it can be tested, for the amount of radiation off $b$ 
quarks relative to light ones at LEP1, the new algorithm indeed is
successful \cite{newshow,QCDWG}.

\begin{figure}[t] 
\centerline{\epsfig{file=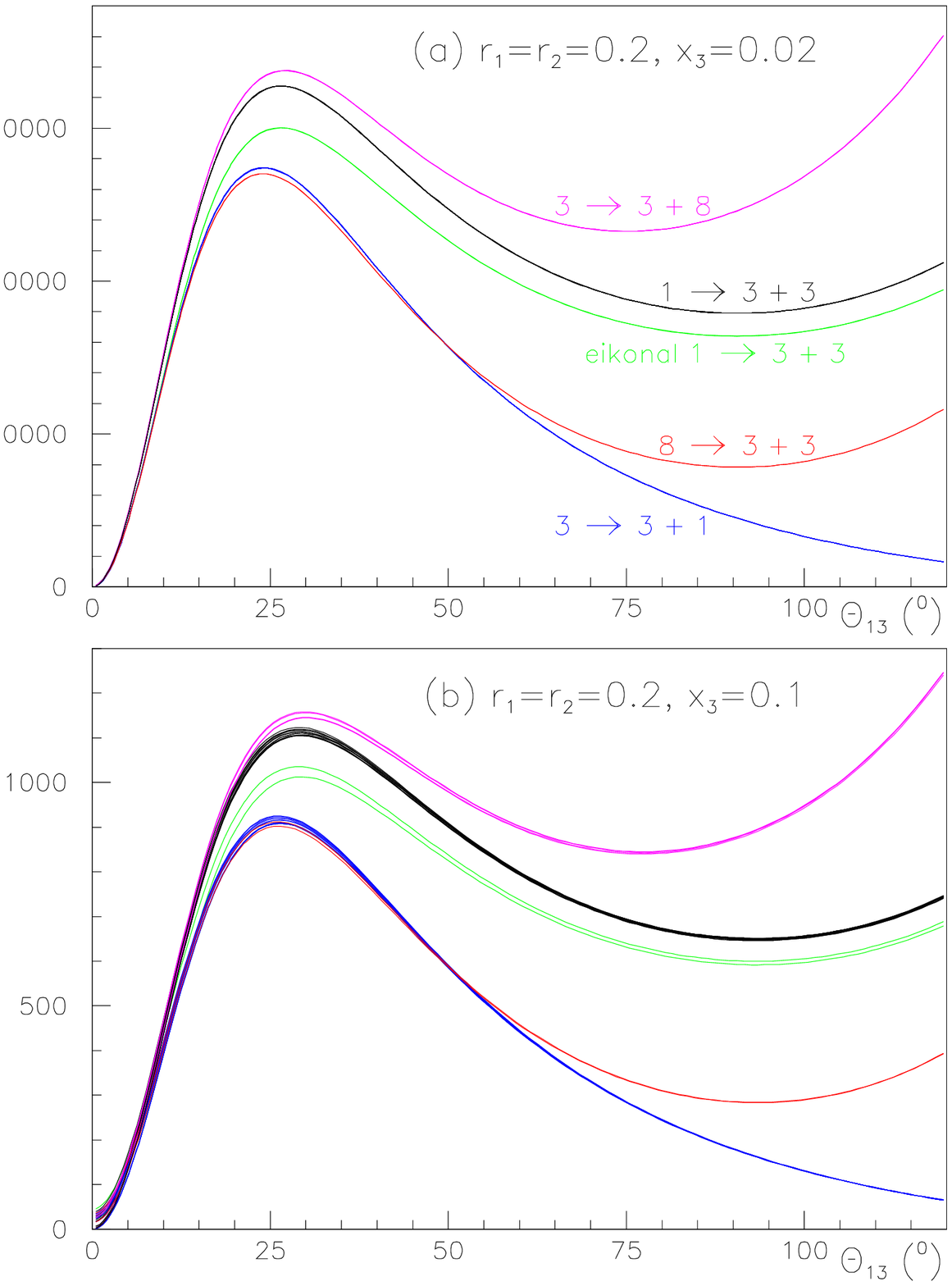,width=8cm}%
\epsfig{file=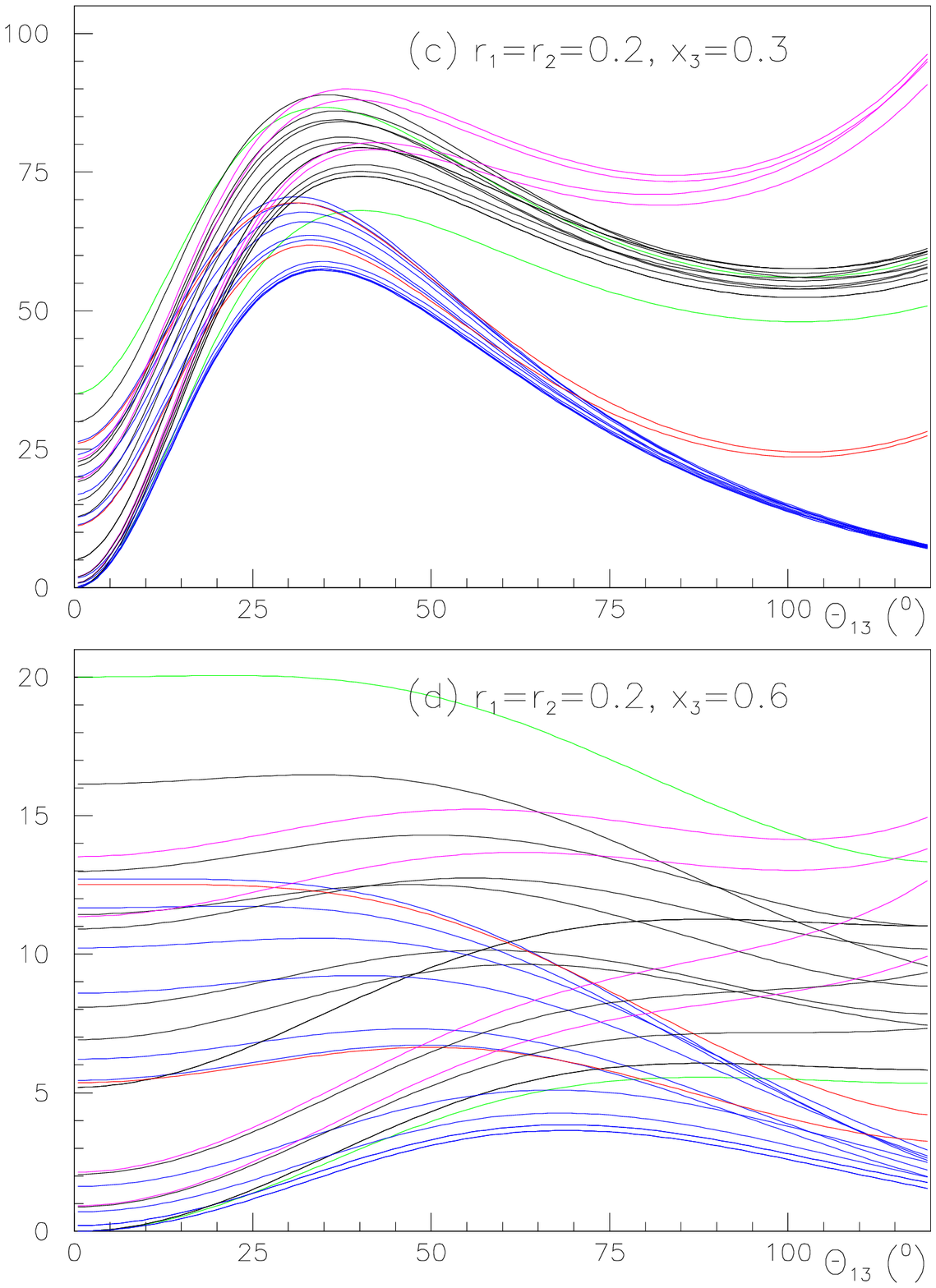,width=8cm}}
\vspace{10pt}
\caption{Gluon radiation pattern $W_{\mathrm{ME}}(x_1,x_2)$, 
eq.~(\protect\ref{WME}), as a function of the gluon emission 
angle, for four fixed gluon energy fractions, 
$x_3 = 2 E_g/E_{\mathrm{CM}}$. The daughters/mother mass ratio is fixed 
at 0.2. The curves are for different combinations of colour and spin,
and with/without a $\gamma_5$ factor where allowed.}
\label{fig1}
\end{figure}

As an illustration of the process dependence, Fig.~\ref{fig1}
shows the radiation pattern of the various matrix elements calculated.
In order to ease the comparison, the same fixed mass ratios have been
used for all processes, $r_1 = r_2 = 0.2$. Furthermore, the large mass
ratio highlights the dead cone effect, which shows a universal
behaviour for small gluon energies. At large angles, and still small 
gluon energies, there is a dependence on the colour structure of the
process, but not e.g. on the spin of the particles. This should be
expected, since in the soft-gluon limit radiation can be described
by a spin-independent eikonal expression \cite{eikonal}. 
Maybe more surprising is how completely this universality
breaks down for more energetic gluons. Then processes are split not only
by colour, but also by the spin structure, and the presence or not 
of a $\gamma_5$ in the matrix element, where allowed. (The figure only 
show the two extremes; by an arbitrary admixture of the two one would
instead obtain a set of allowed bands.) Furthermore, the dead cone
effect is shown to remain only for the case of a spin 0 particle decaying 
to two daughters also with spin 0. In retrospect, the process dependence
is there also at small gluon energies, but is nonsingular and therefore 
invisible underneath the eikonal soft-gluon-singular contributions.

The above figure well illustrates that differences could be big in 
principle, but fortunately the reality is more forgiving. One reason is
the big jump in mass between the $b$ quark, on the one hand, and $t$,
SUSY and any other potentially coloured particles, on the other. The 
most direct consequence is that the heavier particles typically generate
only a small fraction of the total amount of QCD radiation, while
$b$ and lighter quarks produce the bulk of it. The $b$ is light
on the scale of the decaying particle, and so has a smaller dead cone
than the one in Fig.~\ref{fig1}.

A more realistic example of differences is then offered by a light
Higgs state, say 115--130 GeV in mass as suggested by the MSSM scenario, 
decaying to $b\overline{b}$. The three-jet rate in such events typically
is 10--15\% higher than in $\gamma^*/Z^* \to b\overline{b}$ (or light 
quark) decays at the same energy. The difference is less for soft 
radiation, so the Higgs decay is only producing about 1\% lower mean 
values for the $b$ quark and $B$ hadron fragmentation functions.

In $t\overline{t}$ events, the new algorithm increases the amount of
radiation in the top production stage, but decreases it in the subsequent
top decay. The difference is especially notable in the $W$ hemisphere of 
the top decay, where the gluon emission rate is dropping rather steeper
(with the angle away from the $b$ quark) in the new program than in the 
old. This is related to a destructive interference between emission off 
the $t$ and off the $b$ in this hemisphere, while the older approach had 
its origins in $e^+ e^- \to q\overline{q}$ events, where the interference 
is constructive. The net result is a small but visible decrease in the 
total amount of gluon radiation in $t\overline{t}$ events.

For supersymmetric processes, results largely depend on the actual masses.
Assuming the charged Higgs mass to coincide with the $W^{\pm}$ one, the 
decays $t \to b W^+$ and $t \to b H^+$ give almost identical amounts of 
radiation. But if the stop mass agrees with the top one, there is 
more QCD radiation in the former production process than the latter. 
(While the difference in threshold behaviour here gives the opposite effect
for ISR photon radiation, which can become more important.)

\end{document}